\documentclass[prl,twocolumn,showpacs,nofootinbibs,superscriptaddress
]{revtex4}
\usepackage{epsf}
\usepackage{epsfig}
\usepackage{amsmath}

\newcommand{\be}{\begin{eqnarray}}
\newcommand{\ee}{\end{eqnarray}}
\newcommand{\ba}{\begin{array}}
\newcommand{\ea}{\end{array}}
\newcommand{\tr}{\mbox{\rm tr}}

\begin{document}
\preprint{}

\title{Chiral Logarithms Tamed}

\author{Nikolai Kivel}
%\email{}
\affiliation{Petersburg Nuclear Physics
Institute, Gatchina, St.\ Petersburg 188350, Russia}
\affiliation{Institut f\"ur Theoretische Physik II,
Ruhr--Universit\"at Bochum, D--44780 Bochum, Germany}

\author{Maxim V. Polyakov}
%\email{}
\affiliation{Petersburg Nuclear Physics
Institute, Gatchina, St.\ Petersburg 188350, Russia}
\affiliation{Institut f\"ur Theoretische Physik II,
Ruhr--Universit\"at Bochum, D--44780 Bochum, Germany}

\author{Alexei Vladimirov}
%\email{}
\affiliation{Institut f\"ur Theoretische Physik II,
Ruhr--Universit\"at Bochum, D--44780 Bochum, Germany}
\affiliation{Bogolubov Laboratory of Theoretical Physics, JINR, 141980 Dubna, Russia}

\date{\today}

\begin{abstract}
We derive non-linear recursion relations for the leading chiral
logarithms (LLs). These relations not only provide a very
efficient method of computation of LLs (e.g. the  33-loop
contribution is calculated in a dozen of seconds on a PC) but also
equip us with a powerful tool for the summation of the LLs. Our
method is not limited to the chiral perturbation theory only, it
is pertinent for any non-renormalizable effective field theory
such as, for instance, the theory of critical phenomena,
 the low-energy
quantum gravity, etc.
\end{abstract}

\pacs{}

\maketitle
\thispagestyle{empty}

The very fact of the spontaneous breakdown of the approximate
chiral symmetry in strong interactions leads to the possibility of
the systematic expansion of hadronic amplitudes and correlation
functions at low energies. The expansion is organized in powers of
external momenta and the masses of pseudo-Goldstone bosons
(pions). We denote the corresponding expansion parameter
generically as $p^2$. An efficient method to perform the chiral
expansion is based on the technique of effective chiral Lagrangian
\cite{weinberg79}. The leading $O(p^2)$  hadronic amplitudes can
be obtained from the tree diagrams of the famous Weinberg Lagrangian \cite{weinberg68}:

\be
\label{L2}
\mathcal{L}_{2}=\frac{F^{2}}{4}\text{~} \text{tr}\left[ \left(  \partial_\mu U\partial_\mu
U^{\dagger}\right)  +m^{2}\left(  U+U^{\dagger}\right)  \right]\, ,
\ee
where $F$ is the pion decay constant in the chiral limit, and $m$ is the pion mass.  The chiral corrections of the order $O(p^4 \ln p^2)$ can be obtained from the
 one-loop calculation with the Weinberg Lagrangian (\ref{L2}) \cite{Pagels,weinberg79}. In order to compute higher corrections, e.g. $O(p^4)$,
 $O(p^6 \ln p^2)$, etc., one has to include terms with four and higher derivatives in the effective chiral Lagrangian \cite{gasser}.
Note, however, that the leading chiral logarithms (LLs), i.e. the correction of the form $O(p^2[p^2\ln p^2]^n)$ can be obtained
from the $n$-loop diagrams generated by the Weinberg Lagrangian (\ref{L2}). The conciseness and beauty of the leading logarithm approximation lies
in the fact that LL corrections depend only on one basic low-energy constant -- $F$. In this approximation one avoids rapid proliferation of the
low-energy constant with increasing of the chiral order.
The calculation of LLs is a herculean task -- it requires the computation of $n$-loop diagrams in the
{\it non-renormalizable} field theory (\ref{L2}). Presently, the LLs are computed to the two-loop accuracy for the $\pi\pi$-scattering amplitude \cite{colangelo1},
to the five-loop accuracy for the correlator of scalar currents \cite{bissegger}, and to the three-loop accuracy for the generalized parton
distributions (GPDs) \cite{my}. We note that for the case of the chiral corrections to GPDs the summation of LLs is indispensable \cite{my,Kivel08}, because
the smallness of the chiral expansion parameter is compensated by  $1/x_{\rm Bj}^n$.

In this Letter we present an efficient method to compute LLs. This method reduces the task of LLs computation to a simple algebraic problem, and paves a way towards
the summation of LLs.

We discuss details of the method for the
 massless $O(N+1)$ $\sigma$-model defined by the Lagrangian:

\be
\label{ON}
\mathcal{L}_{2}&=&\text{~}\frac{1}{2}\left[  \partial_{\mu}\sigma\partial_{\mu
}\sigma+\partial_{\mu}\pi^{a}\partial_\mu\pi^{a}\right] \, ,
\ee
where the fields are constrained by the relation $\sigma^{2}+\sum_{a=1}^{N}\pi^{a}\pi^{a}=F^{2}\, .$
We consider the $O(N+1)$ $\sigma$-model for the following reasons:
\begin{itemize}
\item
It is equivalent to the massless two-flavour Weinberg Lagrangian
(\ref{L2}) for $N=3$.
\item
It is a free field theory for $N=1$, which can be used as a check of our calculations
\item
It can be solved in the large-$N$ limit,
which provides a check of our calculations and allows us to asses the accuracy of the $1/N$ expansion
without tedious calculations.
\end{itemize}
For simplicity,
we consider LLs for the forward $\pi\pi$ scattering amplitude in the massless $O(N+1)$ $\sigma$-model (\ref{ON}).
The reader can easily apply our method to an observable in a field theory of her/his choice.
The forward amplitude computed at Mandelstam $t=0$ has the form:

%\begin{widetext}
\be
\label{amp}
T^{abcd}(s)=\delta^{ab}\delta^{cd}A(s)+\delta^{cb}\delta^{da}B(s)+\delta^{bd}\delta^{ac}C(s).
\ee
%\end{widetext}
The chiral expansion of the functions $A(s),B(s),C(s)$ has the following structure
\be
\label{struktura}
A(s)&=&(4\pi)^2 S\ \sum_{n=0}^\infty \sum_{k=0}^n A_n^{(k)} S^n\ L^k \, ,\\
\nonumber
B(s)&=&(4\pi)^2 S\ \sum_{n=1}^\infty \sum_{k=0}^n B_n^{(k)} S^n\ L^k ,~~C(s)=A(-s)\, ,
\ee
where we introduce a dimensionless invariant energy $S\equiv s/(4\pi F)^2$ and $L$ denotes the chiral logarithms,
$L\equiv \ln(\mu^2/s)$, $\mu$ is the renormalization scale. The first
$A_0^{(0)}$-term in the expression for $A(s)$
corresponds to the tree contribution to the scattering amplitude
in the $O(N+1)$ $\sigma$-model (\ref{ON}). The other terms are higher chiral order corrections. Our aim consists in the
calculation of the LL coefficients $A_n^{(n)}\equiv A_n$ and $B_n^{(n)}\equiv B_n$
appeared in Eq.~(\ref{struktura}).

 Simple power counting \cite{weinberg79}
shows that the coefficient $A_n^{(k)}$ originates from the $k$-loop diagram with vertices from the chiral Lagrangian
$\mathcal{L}_p$ with the number of derivatives of the pion fields $p\leq 2(n+1-k)$. We see that the
$n$-th order LL coefficient
 receives contribution from the $n$-loop diagrams with vertices generated by the leading Lagrangian (\ref{ON}).
The UV divergencies in a $n$-loop diagram are removed by the subtraction of lower-loop graphs with
insertion of the local counterterms corresponding to the subdivergencies of the original $n$-loop diagram. See detailed
discussion of the structure of the subtractions in Refs.~\cite{colangelo,Kivel08}. The local counterterms
relevant
for our calculations renormalize
the couplings the all-order Lagrangian, which encodes the structure of counterterms:

\be
\label{ct}
\mathcal{L}=-\frac{1}{8} \sum_{n=1}^{\infty}\sum_{j=0\atop \scriptstyle{\rm even}}^{n}
\frac{g_{nj}(\mu)}{(4\pi F)^{2n}}\partial^{2n}P_{j}\left(\frac{\partial_1 \partial_2}{\partial^2} \right).
\ee
Here $P_{j}$ are Legendre polynomials and
a convenient notation
for the operator monomials is introduced:

\begin{widetext}
\be
\partial^{2n}\left(\frac{\partial_1 \partial_2}{\partial^2} \right)^{j}\equiv
\left(\pi^a \overleftrightarrow{\partial}_{\nu_1}\ldots \overleftrightarrow{\partial}_{\nu_{j}}\pi^a\right)
\partial^{2(n-j)}
\left(\pi^b \overleftrightarrow{\partial}_{\nu_1}\ldots \overleftrightarrow{\partial}_{\nu_{j}}\pi^b\right)\,.
\ee
\end{widetext}
The coupling constants $g_{nj}(\mu)$ are enumerated by two indices. The index $n$ indicates the number of
derivatives of the pion fields ( equal to $2n$) in the corresponding counterterm. We refer to the index $n$ as ``principal index".
The second index $j$ corresponds to the ``exchanged spin" of the counterterm.
The tree level contribution of  the vertices (\ref{ct}) to the amplitude can be easily computed with the result:
\be
\label{ampct}
A_{\rm tree}(s)&=&-\sum_{n=1}^\infty (-S)^n\sum_{j=0\atop \scriptstyle{\rm even}}^{n}g_{nj}(\mu)\, ,\\
\nonumber
B_{\rm tree}(s)&=&-\sum_{n=2\atop \scriptstyle{\rm even}}^\infty (S)^{n} g_{n n}(\mu)\ \frac{(2n)!}{n!n!}\, .
\ee
The expansion coefficients $A_n^{(k)}$ and $B_n^{(k)}$ of the amplitude (\ref{struktura}) are functions of the infinite
set of couplings $g_{nj}(\mu)$, denoted  by $\bf g$. These coefficients depend on the renormalization scale $\mu$
through $\mu$-dependence of the couplings $\bf g$. The renormalized (physical) amplitude is given by the sum of
Eq.~(\ref{struktura}) and Eq.~(\ref{ampct}) and it must be independent of $\mu$.
Thus, imposing the requirement that
$\mu^2 \frac{d}{d\mu^2} (A(s)+A_{\rm ct}(s))=0$ we obtain\footnote{Below we discuss the amplitude $A$ only, the discussion
for the amplitudes $B$ and $C$ is analogous. For the latter we provide only the final result} the following set of equations:

\be
\label{uravn}
A_n^{(1)}({\bf g})+(-1)^{n}\sum_{j=0\atop \scriptstyle{\rm even}}^{n+1}
\beta_{n+1j}({\bf g})=0\\
\nonumber \hat{H} A_n^{(k)}({\bf g})+(k+1)A_n^{(k+1)}({\bf
g})=0\,. \ee
Here the $\beta$-functions are defined as
$\beta_{nj}({\bf g})\equiv \mu^2 \frac{d}{d\mu^2} g_{nj}(\mu)$. Also we have introduced
the differential operator $\hat{H}$ acting on the space of the
coupling constants:
\be \label{H}
\hat{H}\equiv \sum_{n=1}^\infty \sum_{j=0\atop \scriptstyle{\rm
even}}^{n}\beta_{nj}({\bf g})\frac{\partial}{\partial g_{nj}}\, .
\ee The set of equations (\ref{uravn}) has the following solution:
 \be
\label{solu} A_n^{(k)}({\bf g})=\frac{(-1)^{n+k}}{k!}\hat{H}^k
\sum_{j=0\atop \scriptstyle{\rm even}}^{n+1} g_{n+1j}. \ee
The
lowest constant $A_0^{(0)}=g_{10}=1$ is fixed by the tree level
calculations with the Lagrangian (\ref{ON}). We see from the
solution (\ref{solu}) that, in order to obtain the LLs (constants
$A_n^{(n)}$), we have to apply the operator $\hat H$ $n$ times to a
linear combination of the coupling constants $g_{nj}$. This at
first glance formidable problem can be solved if one notes the
following crucial property of the operator $\hat{H}$: \be
\label{crucial} \hat{H}^n\ g_{mj}=0,~~ {\rm if}~~m < n\, . \ee
Indeed, the loop diagrams contributing to the renormalization of the
constant $g_{mj}$ include vertices with constants $g_{pl}$ with
$p < m$ only. It implies that the $\beta_{mj}(\bf g)$-functions
depend only on the subset of the low-energy constants $g_{pl}$
with $p< m$. Hence, it is easy to see that the application of
the operator $\hat H$ to $g_{mj}$ leads to the lowering of the
principal index $m$ by one unit, i.e. we can write the general
form of the action of $\hat{H}^n$ on $g_{n+1j}$: \be
\label{uravneniedlyomega} \hat{H}^n g_{n+1j}= n!\ \omega_{nj}
g_{10}^{n+1}= n!\ \omega_{nj}\, . \ee The expression (\ref{uravneniedlyomega}) simultaneously presents the
definition of quantities $\omega_{nj}$ ($n$ is the number of
loops and $j\leq (n+{\rm Mod}(n,2))$),
 which
determine the LL coefficients $A_n\equiv A_n^{(n)}$ (see Eq.~(\ref{solu}) with $k=n$).

Now our aim is to calculate $\omega_{nj}$ from Eq.~(\ref{uravneniedlyomega}). Due to the
property (\ref{crucial}) only one-loop  piece of $\beta_{nj}$-functions, which is quadratic in couplings $g_{nj}$, contribute to Eq.~(\ref{uravneniedlyomega}).
This observation is in accordance with general RG analysis of Ref.~\cite{colangelo}. The general structure of the one-loop
$\beta_{nj}$-function is the following:
%\begin{widetext}
\be
\label{1loop}
\beta_{nj}^{\rm 1-loop}({\bf g})=
\sum_{m=1}^{n-1}\sum_{i=0\atop \scriptstyle{\rm even}}^{m}\sum_{l=0\atop \scriptstyle{\rm even}}^{n-m}
{\mathcal B}_{j}^{(m,i)(n-m,l)}\ g_{mi} g_{(n-m)l}
\ee
Here ${\cal B}_{j}^{(m,i)(n-m,l)}$ are numerical coefficients which can be obtained from the calculation of one-loop diagrams
shown in Fig.~1.
\begin{figure}
\includegraphics[width =7.cm]{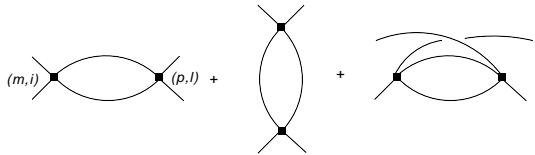}
\caption{One loop diagrams contributing to the $\beta$-functions coefficients (\ref{bety}). Filled squares denote the
counterterms $(mi)$ and $(pl)$ introduced in Eq.~(\ref{ct}). }
\label{fig:halla}
\end{figure}
The result of calculations gives the following result:
\begin{widetext}
\be
\label{bety}
{\mathcal B}_{j}^{(m,i)(p,l)}=\frac{1}{2j+1}\left[\frac{N}{2}\delta_{ij}\delta_{lj}+\delta_{ij} \Omega^{li}_{p}
+\delta_{l j} \Omega^{il}_{m}
\right]+\left(1+(-1)^j\right)\sum_{k=0}^{{\rm min }[p,m]} \frac{\Omega_m^{ik} \Omega_{p}^{lk} \Omega_{m+p}^{kj}}{2 k+1}\, ,
\ee
\end{widetext}
where constants $\Omega_n^{AB}$ are computed as the following integral with Legendre polynomials:
\be
\label{omega}
\Omega_n^{AB}=\frac{2 B+1}{2^{n+1}}\int_{-1}^1 dx P_A\left(\frac{x+3}{x-1}\right) P_B(x) (x-1)^n.
\ee
The $(n+1)\times (n+1)$ matrix $\hat{\Omega}_n$ with matrix elements given by Eq.~(\ref{omega}), posses
many beautiful and intriguing properties. For example,
$\hat{\Omega}_n^2=1$ and $\tr(\hat{\Omega}_n)=\sin(\pi n) $, which follow from the fact that the matrix $\hat{\Omega}$ represents
$SL(2,R)$ transformations.
The relation of this symmetry group to the general structure of the renormalization procedure in a wide class of effective quantum field theories will be discussed elsewhere.

Now substituting the one-loop $\beta$-function (\ref{1loop}) into Eq.~(\ref{uravneniedlyomega}) we obtain
the following nonlinear recursive relations for desired coefficients $\omega_{nj}$:
\be
\label{main}
\omega_{nj}=\frac{1}{n}\sum_{m=0}^{n-1}\sum_{i=0\atop \scriptstyle{\rm even}}^{m+1}\sum_{l=0\atop \scriptstyle{\rm even}}^{n-m}
{\mathcal B}_{j}^{(m+1,i)(n-m,l)} \omega_{mi}\omega_{(n-m-1) l},
\ee
The recursion (\ref{main}) allows  to express the higher coefficients $\omega_{nj}$ through that with lower principal indices, starting with $\omega_{00}=1$.
[We remind that $n$ enumerates the loop order and $j\leq (n+{\rm Mod}(n,2))$]
The coefficients of the $\beta$-functions  ${\mathcal B}_{j}^{(m+1,i)(n-m,l)}$ are given by Eq.~(\ref{bety}). The LLs for the amplitudes $A(s), B(s)$
[($C_n=(-1)^{n+1} A_n$)]
can be computed in terms of $\omega_{nj}$ as follows:
\be
A_n=\sum_{j=0\atop \scriptstyle{\rm even}}^{n+1} \omega_{nj}\,,~~B_n=\frac{(2n+2)!}{[(n+1)!]^2} \omega_{nn+1}
\ee

The recursive relation (\ref{main}) is the main result of the
present paper. It allows a  very fast computation of LLs. For
example, the 33-loop chiral LL is computed in a dozen of seconds
on a PC \footnote{Mathematica notebook for computing LLs is
available at
http://www.tp2.rub.de/$\sim$maximp/research/research.html}. The
6-loop results for LLs  are presented \footnote{We are limited in
the presentation of the results by the page size only.} in Table~A
for the amplitude $A(s)$ and 7-loop results in Table~B
  for the amplitude $B(s)$.

Our results for LLs for $N=3$ agree with  two-loop calculations of
$\pi\pi$ amplitude \cite{colangelo1} and with five-loop results
for the correlator of the scalar currents \cite{bissegger}.
Additional check of our method is provided by the case of $N=1$.
Indeed, for that value of $N$ the Lagrangian (\ref{ON})
corresponds to a free field theory, therefore we should obtain
nullification of LLs. For $N=1$ the scattering amplitude is given
by the sum $A(s)+B(s)+C(s)$, it is easy to see that all LLs are
cancelled in this case.

The non-linear recursion relation (\ref{main}) for LLs is valuable
not only because it renders a breakthrough in the calculations of
LLs, it also allows general theoretical studies of LLs anatomy in
wide class of effective theories. As an example, let us consider
the behaviour of LLs in the large-$N$ limit of $O(N+1)$
$\sigma$-model. In this limit we can neglect all terms in the
expression (\ref{bety}) but the first one. Then, the recursion
relation (\ref{main})  for large $N$ LLs is simplified
considerably: \be \omega_{nj}^{\rm LN}=\frac{1}{2 n
}\sum_{m=1}^{n-1} \frac{N}{2 j+1}\ \omega_{mj}^{\rm
LN}\omega_{(n-m-1)j}^{\rm LN}. \ee The solution of this recursion
is obvious: \be \label{LNLL} \omega_{nj}^{\rm
LN}=\left(\frac{N}{2}\right)^n\ \delta_{j0}. \ee This solution is
in agreement with the direct large-$N$ calculations in the
$O(N+1)$ $\sigma$-model, see e.g. \cite{Coleman}. In order to
compute the $1/N$ corrections to the leading result (\ref{LNLL})
we substitute the LLs in the form: \be
\omega_{nj}=\left(\frac{N}{2}\right)^n\left[ \delta_{j0}+
\frac{c_{nj}^{(1)}}{N}+\ldots\right]\,  \ee into the recursion
relation (\ref{main}) and obtain the linear equation for the
coefficients $c_{nk}^{(1)}$. This equation has a solution in terms
of the Lerch function. The corresponding expression is rather
long, instead we give the result for the LL coefficients $A_n$ for
the leading and subleading $1/N$ orders in the case of large
number of loops $n\gg 1$. The leading $1/N$ asymptotic of the
amplitude $B$ we compute without any assumptions about $n$. The
result is: \be \nonumber A_n &=& \left(\frac{N}{2} \right)^n
\left[1-\left(\frac{\pi^2}{3}-8(1-\ln 2)\right)
\frac{n}{N}+\ldots\right]\, ,\\
\nonumber
B_n&=&\left(\frac{N}{2}\right)^{n-1} \frac{2}{2+n}\left[1+O\left(\frac{1}{N}\right)\right].
 \ee
 It is a remarkable
result! It shows that the $1/N$ expansion for $O(N+1)$
$\sigma$-model fails in the chiral order $n\sim N$ and the
expansion requires reordering.

In summary, we have developed the method of
non-linear recursion relations (\ref{main}) which allows a calculation the leading chiral logarithms to essentially unlimited order.
Furthermore, this method presents a puissant tool
for  study of general structure of infrared logarithms. It can be applied to any physics problem described by a non-renormalizable
effective low-energy Lagrangian, e.g. theory of critical phenomena, low-energy quantum gravity,
theory of magnetics, etc.

\begin{acknowledgments}

This work was supported in parts by the Alexander von Humboldt Foundation, by BMBF,
 by the Deutsche Forschungsgemeinschaft,
 the Heisenberg--Landau Programme grant,
 and the Russian Foundation for Fundamental Research
 grants No.\ 06-02-16215 and 07-02-91557

\end{acknowledgments}

\begin{widetext}
\renewcommand{\arraystretch}{0.2}
\be\label{table1}
\scriptsize
\nonumber
\text{Table A}\\
\scriptsize
\nonumber
\begin{array}{c|l|l}
\hline
\hline
\# \text{ loops} & N=3 & \text{ Arbitrary}~~ N \\
\hline
1 & \frac{2}{3} & \frac{N}{2}\left(1-\frac{5  }{3N}\right) \\
 \hline
2 & \frac{25}{18} & \frac{N^2}{4}\left(1-\frac{37}{18 N}+\frac{49}{18 N^2}\right) \\
 \hline
3 & \frac{577}{540} & \frac{N^3}{8}\left(1-\frac{287}{90 N}+\frac{407}{90 N^2}-\frac{448}{135 N^3}\right)
\\ \hline
4 & \frac{1481}{864} & \frac{N^4}{16}\left(1-\frac{20753}{5400 N}+\frac{363091}{48600 N^2}-\frac{17849}{2430 N^3}+\frac{404}{81 N^4}\right)
\\ \hline
5 & \frac{28943}{19440} &\frac{N^5}{32}\left(1-\frac{12533}{2625 N}+\frac{7655843}{708750 N^2}-\frac{1319666}{91125 N^3}+\frac{38082031}{3189375 N^4}-\frac{2449121}{425250 N^5}\right)
\\ \hline
6 & \frac{33744493}{15876000} & \frac{N^6}{64}\left(1-\frac{3632171}{661500 N}+\frac{3511547989}{238140000 N^2}-\frac{2119277851}{89302500 N^3}+\frac{6141878783}{238140000
   N^4}-\frac{6249981863}{357210000 N^5}+\frac{954322601}{119070000 N^6}\right)
\\ \hline
%7 & \frac{15149979225613}{7561421280000} & \frac{N^7}{128}\left(1-\frac{19619647}{3087000 N}+\frac{151108200961}{7779240000 N^2}-\frac{1932361048417}{52509870000 N^3}+\frac{89462519644649}{1890355320000
 %  N^4}-\frac{40057482076643}{945177660000 N^5}+\frac{532452359791}{21003948000 N^6}-\frac{861566561867}{94517766000 N^7}\right)
%\\ \hline
\end{array}
\ee
\renewcommand{\arraystretch}{0.2}
\be\label{table2}
\scriptsize
\nonumber
\text{Table B}\\
\scriptsize
\nonumber
\begin{array}{c|l|l}
\hline
\hline
\# \text{ loops} & { N=3} & \text{ Arbitrary}~~ N \\
\hline
1 & \frac{2}{3} & \frac{2}{3} \\
\hline
3 & \frac{181}{270} & \frac{N^2}{10}-\frac{203 N}{1080}+\frac{361}{1080} \\
 \hline
5 & \frac{3320747}{5103000} & \frac{N^4}{56}-\frac{8609 N^3}{126000}+\frac{892579 N^2}{5670000}-\frac{17367061 N}{102060000}+\frac{2914931}{20412000} \\
 \hline
7 & \frac{2433747559349}{3780710640000} & \frac{N^6}{288}-\frac{209635 N^5}{10668672}+\frac{266068577671 N^4}{4480842240000}-\frac{1616650409063
   N^3}{15122842560000}+\frac{16393400307287 N^2}{120982740480000}-\frac{179751202247 N}{1728324864000}+\frac{3414208728293}{60491370240000}
\\ \hline
\end{array}
\ee
\end{widetext}

\end{document}